\documentclass{article}

\usepackage{arxiv}
\usepackage[utf8]{inputenc} 
\usepackage[T1]{fontenc}    
\usepackage{hyperref}       
\usepackage{url}            
\usepackage{booktabs}       
\usepackage{amsfonts}       
\usepackage{nicefrac}       
\usepackage{microtype}      
\usepackage{multirow}       
\usepackage[numbers]{natbib}         
\usepackage{tabularx}
\usepackage{lipsum}
\usepackage{graphicx}
\usepackage{lineno}
\usepackage{wrapfig}

\title{Machine learning modeling of the atomic structure and physical properties of alkali and alkaline-earth aluminosilicate glasses and melts}

\author{
 Charles Le Losq \\
  Université Paris Cité\\
  Institut de physique du globe de Paris\\
  CNRS-UMR 7154 \\
  F-75005 Paris, France \\
  \texttt{Corresponding author: lelosq@ipgp.fr} \\
   \And
 Barbara Baldoni \\
  Université Paris Cité\\
  Institut de physique du globe de Paris\\
  CNRS-UMR 7154 \\
  F-75005 Paris, France \\
  \texttt{baldoni@ipgp.fr} \\
}

\begin{document}
\maketitle
\begin{abstract}
The first version of the machine learning greybox model i-Melt was trained to predict  latent and observed properties of K$_2$O-Na$_2$O-Al$_2$O$_3$-SiO$_2$ melts and glasses. Here, we extend the model compositional range, which now allows accurate predictions of properties for glass-forming melts in the CaO-MgO-K$_2$O-Na$_2$O-Al$_2$O$_3$-SiO$_2$ system, including melt viscosity (accuracy equal or better than 0.4 log$_{10}$ Pa$\cdot$s in the 10$^{-1}$-10$^{15}$ log$_{10}$ Pa$\cdot$s range), configurational entropy at glass transition ($\leq$ 1 J mol$^{-1}$ K$^{-1}$), liquidus ($\leq$ 60 K) and glass transition ($\leq$ 16 K) temperatures, heat capacity ($\leq$ 3 \%) as well as glass density ($\leq$ 0.02 g cm$^{-3}$), optical refractive index ($\leq$ 0.006), Abbe number ($\leq$ 4), elastic modulus ($\leq$ 6 GPa), coefficient of thermal expansion ($\leq$ 1.1 10$^{-6}$ K$^{-1}$) and Raman spectra ($\leq$ 25 \%). Uncertainties on predictions also are now provided. The model offers new possibilities to explore how melt/glass properties change with composition and atomic structure.
\end{abstract}

\keywords{glass \and melt \and machine learning \and properties \and viscosity \and density \and aluminosilicates}

\section{Introduction}
Aluminosilicate melts that contain alkali and calc-alkaline metal
cations serve as the base composition in the glass-making industry and
also constitute the liquid fraction of most of the Earth's magmas. The
viscosity of these melts is critical as it determines their resistance
to movement at high temperatures, thereby influencing their mobility.
This property heavily impacts the fragmentation of magmas in volcanic
edifices (e.g., see reviews of \cite{gonnermann2013, gonnermann2015}), and working temperatures in
industrial glass-making furnaces. Other properties of melts and glasses,
such as density or optical refractive index, may also be of significant
interest because they can influence not only the mobility of the melt,
for example through buoyancy effects, but also the weight and optical
properties of glass objects. The prediction of such physical properties
is, therefore, essential in addressing problems ranging from the
dynamics of volcanic eruptions to the development of novel glass
materials.

Viscosity predictions can be made using empirical  \cite{bottinga1972, shaw1972, persikov1991, hui2007, giordano2008, duan2014} or
thermodynamic models \cite{sehlke2016,lelosq2017,starodub2019}.  These models offer direct and convenient estimations of viscosity for specific compositional systems,  rendering them highly specialized.  Besides, empirical models rely on predetermined functions, while thermodynamic models depend on our limited understanding of the thermodynamics of silicate melts.  As a result, no universal viscosity model has been developed to predict viscosity for a wide range of temperatures and compositions in the field of glass-making and volcanic silicate melts.  Alternatively,  molecular dynamics (MD) simulations provide an alternative approach to overcome these limitations, offering valuable property predictions for glasses and melts  \citep[e.g., ][]{guillot2007a, spera2011,karki2013,bauchy2013, wang2014, dufils, bajgain2021,bajgain2022}. These simulations are particularly useful as they can provide data for conditions that are challenging to replicate in experiments, such as pressures relevant to planetary magma oceans \cite[e.g., ][]{bajgain2022} or for "exotic" compositions like pure MgO \cite{bogels2022}.  However,  MD predictions are typically limited to relatively high temperatures due to computational costs. Consequently,  it remains extremely challenging to systematically study melt properties across a wide temperature range, spanning from supercooled to superliquidus temperatures, for hundreds or even thousands of compositions at the present time."

To circumvent such limitations, a new set of models relying on
machine learning have been proposed: greybox models. Those combine
physical/thermodynamic equations with machine learning to predict
melt/glass property, with good success to date \cite{tandia2019,liu2019a, hwang2020, lelosq2021}.
Among published greybox models, i-Melt \cite{lelosq2021,lelosq2021d} is a multitask model
that predicts not only melt viscosity through five different equations
but also glass density, optical refractive index and Raman spectrum. It
further provides access to latent properties such as melt fragility or
configurational entropy at the glass transition. i-Melt thus allows the
systematic exploration of the links between composition, structure
(through Raman spectra predictions), and properties of melts and
glasses. The downside of this model is that it currently is limited to
the glass-forming domain of the
Na\textsubscript{2}O-K\textsubscript{2}O-Al\textsubscript{2}O\textsubscript{3}-SiO\textsubscript{2} quaternary system.

In this study, we present a new version of i-Melt that now includes CaO
and MgO. i-Melt was trained on melt and glass compositions in the
Na\textsubscript{2}O-K\textsubscript{2}O-MgO-CaO-Al\textsubscript{2}O\textsubscript{3}-SiO\textsubscript{2} system,
for which a fairly complete, albeit sparse, experimental dataset is
available. In addition to the properties initially predicted by the
first version of the model, it also now predicts melt liquidus
temperatures and heat capacities, as well as glass coefficients of
thermal expansion, elastic modulus, and Abbe numbers. In this
communication, we present the new dataset, the improvements and the
performance of the updated i-Melt model, and we discuss its
possibilities and limits.

\section{Methods}
\label{sec:headings}

\subsection{Datasets and data preparation}

The original database of i-Melt was completed by collecting existing
Raman spectra, optical refractive index, density, Abbe number, elastic
modulus, coefficient of thermal expansion (CTE) of glasses and liquid
heat capacity, viscosity and liquidus temperature of melts in the
Na\textsubscript{2}O-K\textsubscript{2}O-MgO-CaO-Al\textsubscript{2}O\textsubscript{3}-SiO\textsubscript{2} system.
The data were selected via a review of the existing literature as well
as of the SciGlass database, available at
\url{https://github.com/epam/SciGlass}. Melt viscosity, and glass density,
Raman spectra and optical refractive index were selected by hand
following a review of the literature. Abbe number, elastic modulus, CTE
and liquidus temperature data in the
Na\textsubscript{2}O-K\textsubscript{2}O-MgO-CaO-Al\textsubscript{2}O\textsubscript{3}-SiO\textsubscript{2} system
were extracted from the SciGlass database thanks to the GlassPy python
package (version 0.3, \cite{cassar2020}). The data used to train the model as well
as the associated references are provided in the database available in
the Github software repository that hosts i-Melt and its online
calculator (\url{https://github.com/charlesll/i-Melt}), as well as on
Zenodo \cite{lelosq2023b}.

The different streams of data are:

\begin{itemize}
\item \emph{D\textsubscript{viscosity}} (n = 790 compositions), the dataset
of viscosity measurements, composed of
\emph{X\textsubscript{viscosity}} chemical composition entries (mole
fractions) as well as their associated temperatures (Kelvin) and
\emph{y\textsubscript{viscosity}} observations (log\textsubscript{10}
Pa·s);

\item \emph{D\textsubscript{density}} (n = 668 compositions), the dataset of
glass density measurements, composed of
\emph{X\textsubscript{density}} chemical composition entries (mole
fractions) and \emph{y\textsubscript{density}} observations (g
cm\textsuperscript{-3});

\item \emph{D\textsubscript{Raman}} (n = 252 compositions), the dataset of
glass Raman spectra, composed of \emph{X\textsubscript{Raman}} chemical
composition entries (mole fractions) and
\emph{y\textsubscript{Raman}} spectra observations (normalised Raman
intensities);

\item \emph{D\textsubscript{optical}} (n = 610 compositions), the 
dataset of glass optical refractive index, composed
of \emph{X\textsubscript{optical}} chemical composition entries (mole
fractions) as well as their associated wavelength ($\mu$m)
and \emph{y\textsubscript{refractive index}} observations;

\item \emph{D\textsubscript{Cpl}} (n = 95 compositions), the dataset
of liquid heat
capacities $C_{p}^{liquid}$, composed
of \emph{X\textsubscript{Cpl}} chemical composition entries (mole
fractions) as well as their associated temperatures (Kelvin)
and \emph{y\textsubscript{Cpl}} observations (J
mol\textsuperscript{-1} K\textsuperscript{-1});

\item \emph{D\textsubscript{Abbe}} (n = 296 compositions), the  dataset
of glass Abbe Number, composed of \emph{X\textsubscript{Abbe}} chemical
composition entries (mole fractions) as well as their associated Abbe
Numbers (no unit);

\item \emph{D\textsubscript{elastic}} (n = 1006 compositions), the 
dataset of glass elastic modulus, composed of
\emph{X\textsubscript{elastic}} chemical composition entries (mole
fractions) as well as their associated elastic modulus (GPa);

\item \emph{D\textsubscript{CTE}} (n = 2122 compositions), the  dataset
of glass coefficients of thermal expansion, composed
of \emph{X\textsubscript{CTE}} chemical composition entries (mole
fractions) as well as their associated coefficients of thermal expansion
(K\textsuperscript{-1});

\item \emph{D\textsubscript{liquidus}} (n = 4505 compositions), the dataset
of liquidus temperatures, composed of
\emph{X\textsubscript{liquidus}} chemical composition entries (mole
fractions) as well as their associated liquidus temperatures (K).
\end{itemize}

\begin{figure}
    \centering
    \includegraphics[width=\textwidth]{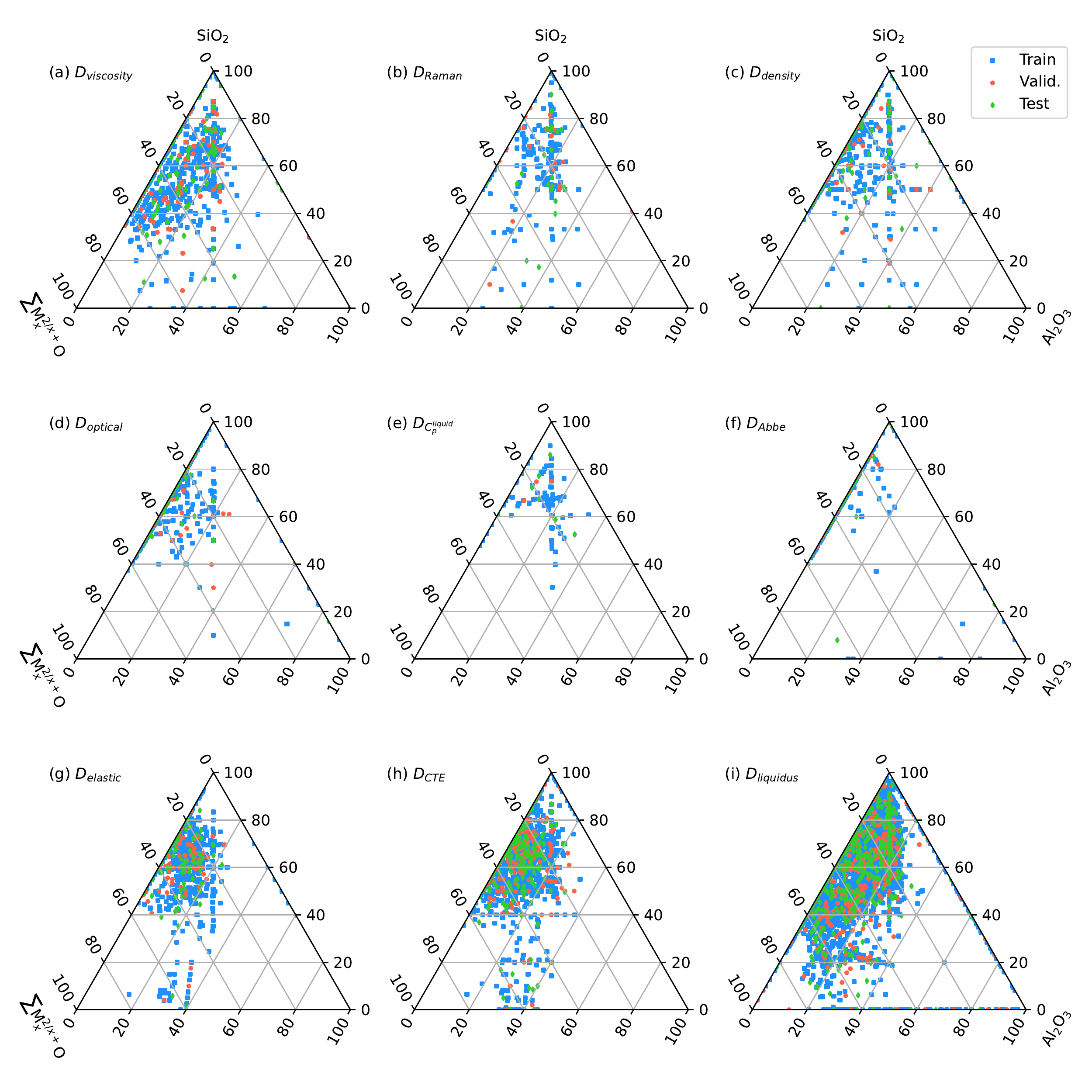}
    \caption{Datasets for melt viscosity (a), glass Raman spectroscopy (b), glass density (c), glass refractive index (d), liquid heat capacity (e), glass Abbe Number (f), glass Elastic Modulus (g), glass coefficient of thermal expansion (h), and liquidus temperature (i). Each symbol corresponds to a sample.}
    \label{fig:fig1}
\end{figure}

The size of the \emph{D\textsubscript{viscosity}}
, \emph{D\textsubscript{density}}, \emph{D\textsubscript{optical}}
, \emph{D\textsubscript{CTE}},
 \emph{D\textsubscript{elastic}} and \emph{D\textsubscript{liquidus}} datasets
allow training i-Melt with a \emph{``high performance''} mindset, because
those datasets cover an important part of the glass-forming domain of
alkali and alkaline-earth aluminosilicates (\textbf{Fig. \ref{fig:fig1}}). The
liquid heat capacity dataset \emph{D\textsubscript{Cpl}} is small
(\textbf{Fig. \ref{fig:fig1}e}), but actually this is not a problem because good
fits of viscosity data with the Adam-Gibbs theory already require the
prediction of consistent liquid heat capacities (see \emph{Results}
section). However, having a \emph{D\textsubscript{Cpl}} dataset, even
small, allowed fine-tuning the model and ensuring
that $C_{p}^{liquid}$ predictions are
consistent with
existing $C_{p}^{liquid}$
data. \emph{D\textsubscript{Abbe}} also is fairly limited (\textbf{Fig.
\ref{fig:fig1}f}), so non-negligible errors on Abbe number predictions may be
expected. Regarding \emph{D\textsubscript{Raman}}, it also covers a
limited set of compositions (\textbf{Fig. \ref{fig:fig1}b} ) and we also do not
expect a very high precision on Raman spectra predictions. Raman data
were actually used as a way of improving learning through a multitask
approach \cite{caruana1997}, because they encode structural information that could
assist the network in learning physical properties, embedding a shared
representation of the composition-structure-property links in melts and
glasses. This actually is one important basis underlying the i-Melt
model: the use of datasets of different glass/melt properties should
allow leveraging their different compositional coverage to ensure better
predictive performance of the model overall.

\begin{wrapfigure}{l}{0.5\textwidth}
    \centering
    \includegraphics{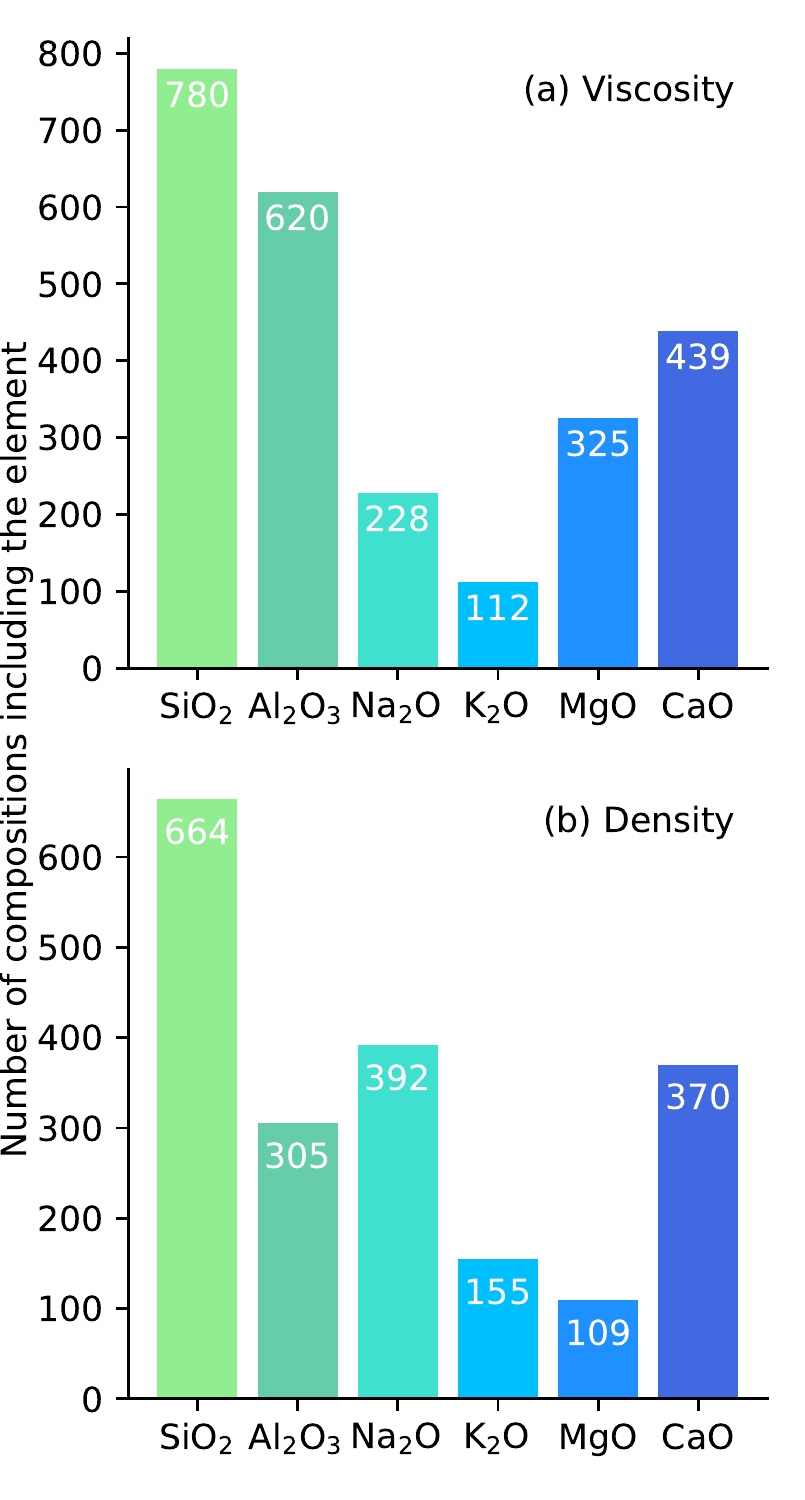}
    \caption{Examples of the numbers of compositions including the different oxide components in the viscosity and density datasets.}
    \label{fig:fig2}
\end{wrapfigure}

Following \cite{lelosq2021} and prior to training i-Melt, the datasets were split
by composition into three different, randomly chosen \emph{training},
\emph{validation} and \emph{testing} subsets. While the \emph{training}
subsets are used for training the model (i.e. tuning its internal
parameters), the \emph{validation} subsets are used for monitoring
overfit and triggering early stopping when the latter occurs \cite{goodfellow2016}.
The final predictive abilities of the trained neural networks are
evaluated using the \emph{testing} data subsets. For performing the
splits, we used a stratified group splitting approach. This approach was
implemented because (i) we need to split the datasets by compositions to
avoid data leakage \cite{kaufman2012}, and (ii) we have imbalanced datasets. While
the former point was taken into account in the original version of the
model \cite{lelosq2021}, the latter was not and becomes particularly important
now. For instance, we have significantly less compositions including
Na\textsubscript{2}O and K\textsubscript{2}O than other elements in the
viscosity dataset, while the density dataset includes less MgO-bearing
compositions (\textbf{Fig. \ref{fig:fig2}}). A train-valid-test split by composition
will work, but may not retain the proportions of each kind of
composition (e.g. sodium silicate, magnesium aluminosilicate, etc.) in
the different train-valid-test splits. To solve this problem, we
assigned to each type of compositions a class (e.g., for sodium silicate
class 1, for sodium aluminosilicates class 2, etc.), and used a
stratified splitting approach that aims at retaining, as much as
possible, the proportions of each class in the different
train-valid-test data subsets. In practice, this approach is implemented
via a hack of the \emph{StratifiedGroupKFold} function of the
scikit-learn library version 1.1.2 \cite{pedregosa2011}. After data splitting, we
systematically checked that there was no sign of data leakage 
(compositions in the train, valid and test subsets show differences
larger than 0.1 mol\%), and we visually checked that the coverage of the
different splits was reasonable (\textbf{Fig. \ref{fig:fig1}}). Train-valid-test
splits were performed with 0.8/0.1/0.1 ratios for all datasets. Scaling of the datasets was done as
described in \cite{lelosq2021}. After pre-processing, the different data subsets
were saved in Hierarchical Data Format HDF5 files for their future use.

\subsection{Machine learning model}

The model i-Melt, implemented in the Python programming language using
the Pytorch machine learning library \cite{paszke2019}, was
presented in detail in \cite{lelosq2021}. We refer the reader to
this publication for an extensive presentation of the model. Here, we
briefly present the model and its general architecture, and focus on
describing improvements and new features.

i-Melt combines an artificial neural network with various dynamic
and thermodynamic equations to predict latent and observed melt/glass
properties. The model takes six inputs: the mole fractions of
SiO\textsubscript{2}, Al\textsubscript{2}O\textsubscript{3},
Na\textsubscript{2}O, K\textsubscript{2}O, MgO and CaO. From these, new
chemical descriptors are now calculated, such as the glass optical
basicity and NBO/T, the Al/M ratio (with M the sum of metal cations),
and the ratio of each element to another. A total of 39 descriptors,
including initial melt composition, are fed into a neural network
composed of \emph{n} hidden layers, each one having \emph{k} activation
units (a.k.a. neurons). The outputs of the hidden layers are fed into
two different linear layers for outputs: the first one returns vectors
that are Raman spectra, and the second one returns 34 different values:

\begin{itemize}
    \item the parameters for the calculation of melt viscosity through five
different theoretical and empirical equations, including Adam-Gibbs \cite{adam1965,richet1984c}, MYEGA \cite{mauro2009},
Avramov-Milchev \cite{avramov1988}, Tamman-Vogel-Fulcher
\cite{fulcher1925} and Free Volume Theory \cite{cohen1979,cohen1984}; 
    \item the partial molar volumes of each oxide components, for density
calculations; 
    \item the partial liquid heat capacity of each oxide component
as well as two temperature-dependent terms for
Al\textsubscript{2}O\textsubscript{3} and K\textsubscript{2}O (following
\cite{richet1985}) for
$C_{p}^{liquid}$ calculations; 
    \item the coefficients of the Sellmeier equation for optical refractive index calculation; 
    \item the melt liquidus temperature; 
    \item the glass Abbe number; 
    \item the glass elastic modulus; 
    \item and the coefficient of thermal expansion of the glass.
\end{itemize}

The artificial neural network allows us, therefore, to input chemical
compositions and obtain predictions for:

\begin{itemize}
\item melt viscosity, within five distinct theoretical or empirical frameworks,
\item melt heat capacity, including partial molar contributions of oxide components,
\item liquidus and glass transition temperatures,
\item latent variables such as configurational entropy and fragility,
\item glass density, including partial molar contributions of each oxide component,
\item glass refractive index as a function of wavelength, its Abbe number and elastic modulus,
\item and the glass Raman spectra.
\end{itemize}

The predictions depend on a large number of adjustable parameters
integral to the neural network, as well as on the careful adjustment of
the neural network hyper-parameters. Adjustment of model parameters
(weights and bias of the activation units) was performed via batch
training through gradient descent using the ADAM optimizer. The global
loss function was calculated from a weighted sum of the root-mean square
errors (RMSE) between measurements and predictions for viscosity as well
as liquid heat capacity, density, optical refractive index, Raman
spectra, liquidus temperature, elastic modulus, CTE, Abbe number and
known glass configurational
entropy $S^{conf}(T_{g})$ values. The
weights assigned to the different tasks in the global loss function were
learned during the optimization process, following the method proposed
by \cite{kendall2018}. Back-propagation was performed using the
automatic differentiation methods implemented in Pytorch \cite{paszke2019}.

Good predictive performance of the model can only be achieved upon
finding optimal sets of model hyperparameters, including the optimizer
learning rate, and the number of layers, the number of units per layer,
and the type of activation functions in the artificial neural network.
Regarding the activation functions, the initial version of i-Melt relied
on Rectifier linear units (a.k.a. ReLU), but new tests showed that
Gaussian error linear units (a.k.a. GELU, see \cite{hendrycks2020}) yield better generalization performance (less overfitting and
better estimates on unseen samples). Therefore, the new version of
i-Melt uses GELU units. To further prevent overfitting and help model
generalization, we also rely on early stopping \cite{goodfellow2016}
and dropout \cite{srivastava2014}. The dropout rate, the optimizer
learning rate, the number of layers and that of activation units per
layer in the artificial neural network were tuned using the
hyperparameter optimization framework Ray Tune \cite{liaw2018},
which allows distributed model selection and training. In practice, we
used the Optuna algorithm \cite{akiba2019} that relied on monitoring
the global loss on the validation datasets to guide the selection of the
most promising models.

This approach allows obtaining a sample of trained models from which we
can select the best ones. Typically, the architecture of the best models
is slightly deep, with 3 to 4 layers, each containing 350 to 500
activation units. The use of GELU units allowed obtaining good model
predictions with dropout rates of \textasciitilde{}0.3. Best ADAM learning
rates were generally in the $1\times10^{-4}$ - $3\times10^{-4}$ range. 
For final predictions, we
average predictions of an ensemble of the 10 best models. As the dropout
rates in the new model are generally high, uncertainties on model
predictions can be estimated using the MC Dropout method \cite{gal2016}: 
for a given input, we leave the dropout active and
ask for, e.g., 100 different samples for each one of the 10 neural
networks. We thus obtain for each input a subset of 1000 predictions,
each prediction being slightly different from the others because coming
from a different neural network, or from a different part of a given
neural network. From this subset, we can provide median values and
confidence intervals, approximating a Bayesian sampling of the model
posterior probability function \cite{gal2016}.

\section{Results}

\subsection{Predictive performance evaluation}

\begin{figure}[ht]
    \centering
    \includegraphics[width=\textwidth]{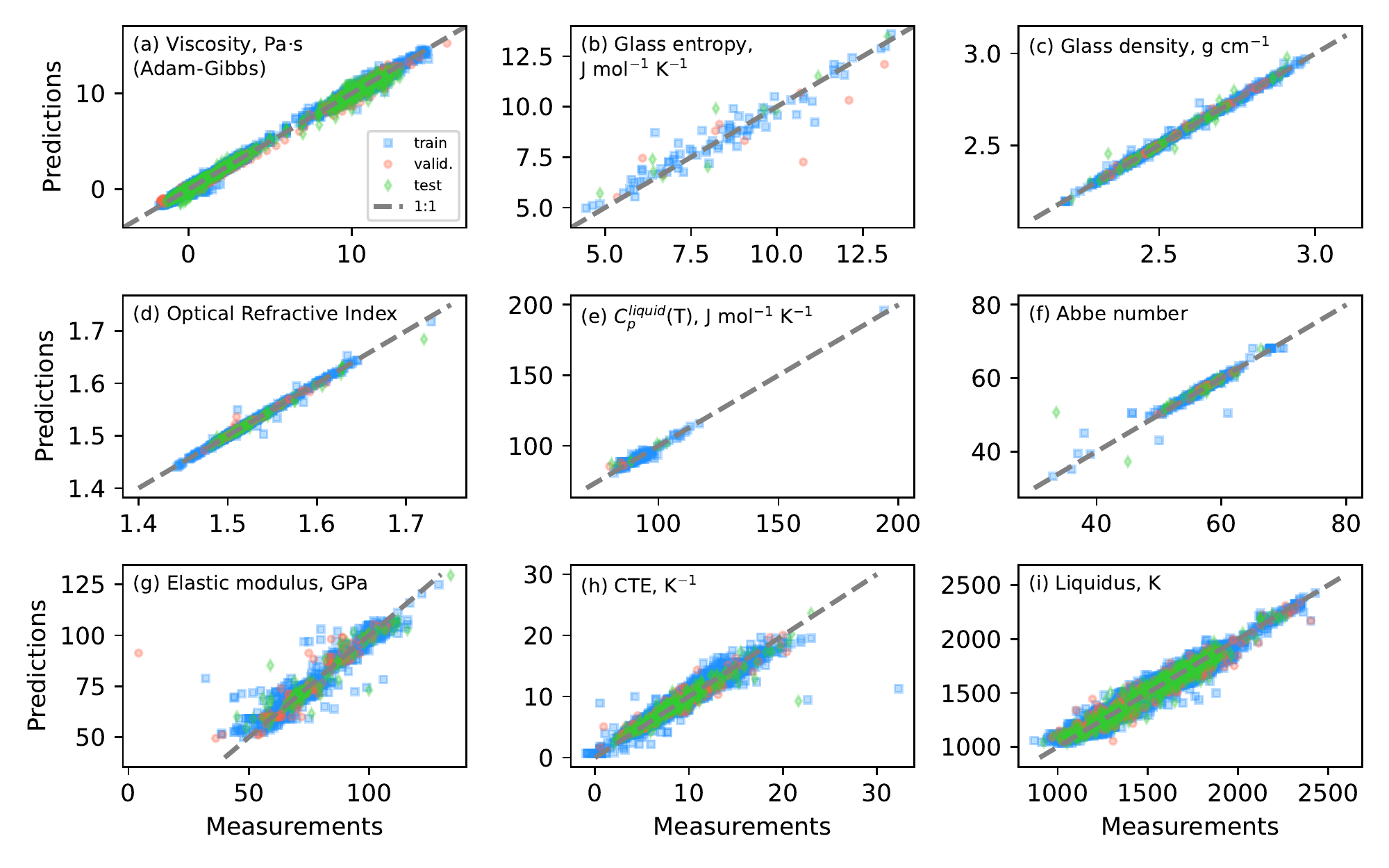}
    \caption{Predictions against measured values for different available observations. Blue squares, orange circles and green diamonds are used for distinguishing the train, validation and testing data subsets.}
    \label{fig:errors}
\end{figure}

Over the very broad range of compositions we investigate (\textbf{Fig.
\ref{fig:fig1}}), viscosity of melts are predicted with a good precision
(\textbf{Fig. \ref{fig:errors}a}). Regardless of the chosen theory, the root mean
squared error (RMSE) values are lower than or equal to 0.4 log\textsubscript{10} Pa·s
on the testing data subset (\textbf{Table \ref{tab:tab1}}). All the median absolute
error (MAE) values, a metric more robust to outliers than RMSE, are
equal to \textasciitilde{}0.1 log\textsubscript{10} Pa·s. In details, predictions in the supercooled
temperature domain are affected by larger errors than predictions in the
sub-liquidus to super-liquidus domain: testing RMSE values (all
equations considered) are in the range 0.4-0.5 log\textsubscript{10} Pa·s
when considering only data in the
10\textsuperscript{7}-10\textsuperscript{15} Pa·s range, whereas they
are in the range 0.2-0.3 log\textsubscript{10} Pa·s for data below
10\textsuperscript{7} Pa·s (\textbf{Table \ref{tab:tab1}}). No significant differences are observed when comparing the predictive errors of the different equations and theories (\textbf{Table \ref{tab:tab1}}). 

\begin{figure}
    \centering
    \includegraphics{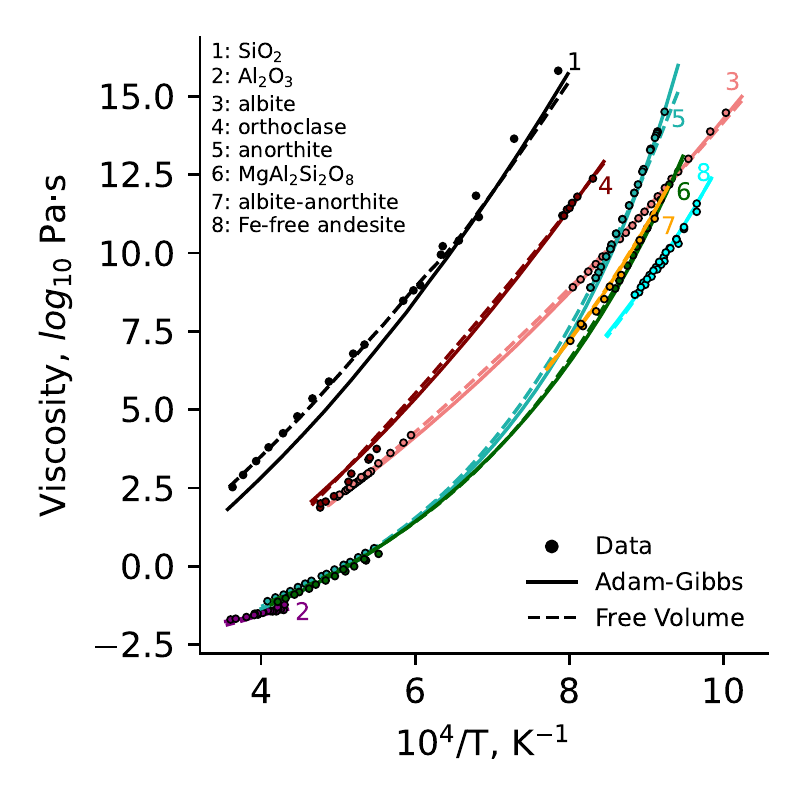}
    \caption{Predictions of the viscosity of selected glass compositions using the Adam-Gibbs and Free Volume theories. Symbols are experimental values from i-Melt database and lines are model predictions using two different theoretical frameworks.}
    \label{fig:fig_visco}
\end{figure}

In \textbf{figure \ref{fig:fig_visco}},  we further show viscosity predictions for specific compositions of interest for geology and industry, for which highly accurate viscosity data are available.  For a given composition, model predictions compare very well with experimental data, regardless of melt compositional complexity. Indeed,  the model predicts very well the viscosity of simple melts such as silica or alumina, but also of  melts in the ternary and quaternary alkali and/or alkaline-earth aluminosilicate systems such as albite, orthoclase, anorthite or its magnesian equivalent.  The viscosity of melts with compositions containing all oxides is also well predicted, as shown for example in  \textbf{figure \ref{fig:fig_visco}} for a melt with an analogue Fe-free andesitic geologic composition.

Overall, the present model uncertainties on viscosity are comparable with, or lower than those of
the previous version of i-Melt for alkali aluminosilicate melts
(\textasciitilde{}0.4 log\textsubscript{10} Pa·s\textbf{)} .  They also
are lower than those affecting existing thermodynamic models for
quaternary alkali aluminosilicate melts \cite{starodub2019,neuville2022a}, or than those affecting empirical models such as
that of Russell and Giordano \cite{russell2005} for albite-anorthite-diopside melts.  The model RMSE on viscosity is lower than that of the more generalistic ViscNet machine learning model of melt viscosity \citep[1.1 on its testing dataset, see][]{cassar2021}.
The accuracy on viscosity predictions of the model actually approaches,
despite a much broader compositional dataset, that of the thermodynamic
model of the viscosity of alkali silicate melts of Le Losq and Neuville
\cite{lelosq2017} (\textasciitilde{}0.2 log\textsubscript{10} Pa·s).

\begin{table}
\centering
\begin{tabularx}{\textwidth}{ X|l|l|l|l} 
\toprule
Data subset & Metric & Training & Validation & Testing\\
\midrule
Viscosity, log\textsubscript{10} Pa·s (Adam-Gibbs theory) & RMSE & 0.2 & 0.3 & 0.4\\
& MAE & 0.1 & 0.1 & 0.1\\
\midrule
Viscosity, log\textsubscript{10} Pa·s (Free Volume theory) & RMSE & 0.2 & 0.2 & 0.3\\
& MAE & 0.1 & 0.1 & 0.1\\
\midrule
Viscosity, log\textsubscript{10} Pa·s (Vogel-Tamman-Fulcher equation) & RMSE & 0.2 & 0.3 & 0.3\\
& MAE & 0.1 & 0.1 & 0.1\\
\midrule
Viscosity, log\textsubscript{10} Pa·s (MYEGA equation) & RMSE & 0.2 & 0.3 & 0.3\\
& MAE & 0.1 & 0.1 & 0.1\\
\midrule
Viscosity, log\textsubscript{10} Pa·s (Avramov-Milchev theory) & RMSE & 0.2 & 0.3 & 0.3\\
& MAE & 0.1 & 0.1 & 0.1\\
\midrule
Density, g cm\textsuperscript{-3} & RMSE & 0.01 & 0.01 & 0.02\\
& MAE & 0.004 & 0.006 & 0.006\\
\midrule
Raman spectra (\%, Median Absolute Percentage Error) & MAPE & 17 & 17 & 25\\
\midrule
Refractive index & RMSE & 0.003 & 0.005 & 0.006\\
& MAE & 0.0009 & 0.0009 & 0.0013\\
\midrule
CTE, 10\textsuperscript{-6} K\textsuperscript{-1} & RMSE & 1.0 & 0.9 & 1.1\\
& MAE & 0.4 & 0.5 & 0.5\\
\midrule
Elastic Modulus, GPa & RMSE & 4 & 8 & 6\\
& MAE & 2 & 2 & 2\\
\midrule
Abbe number & RMSE & 1.1 & 0.4 & 3.7\\
& MAE & 0.3 & 0.2 & 0.5\\
\midrule
Liquidus temperature, K & RMSE & 61 & 66 & 60\\
& MAE & 38 & 39 & 39\\
\midrule
Glass transition temperature (at 10$^{12}$ Pa·s), K & RMSE & 13 & 16 & 12\\
& MAE & 5 & 6 & 9\\
\midrule
Heat capacity, J mol\textsuperscript{-1} K\textsuperscript{-1} & RMSE & 2 & 3 & 3\\
& MAE & 1 & 3 & 1\\
\midrule
Glass entropy,  J mol\textsuperscript{-1} K\textsuperscript{-1} & RMSE & 0.5 & 1.4 & 0.8\\
& MAE & 0.3 & 0.8 & 0.3\\
\bottomrule
\end{tabularx}
\caption{Root Mean Square Error (RMSE) and Median
Absolute Error (MAE) between predictions and measurements. For Raman
spectra, the Median of Absolute Percentage Error (MAPE) is reported. The MAE is
more robust to outliers than the RMSE. Their comparison provides
information regarding the existence of outliers in the datasets, and how
they affect the RMSE.}
\label{tab:tab1}
\end{table}

Other melt/glass properties are also well predicted by i-Melt
(\textbf{Fig. \ref{fig:errors} and Table \ref{tab:tab1}}). In general, training and validation error metrics are very similar, indicating that the model does not overfit. Known viscous glass transition
temperatures $T_g$ and configurational entropy
at $T_g$, $S^{conf}(T_g)$, are predicted
within 16 K and 1 J mol\textsuperscript{-1} K\textsuperscript{-1},
respectively. For $T_g$, such an accuracy is better
than that achieved by the first version of i-Melt (19 K), while
for $S^{conf}(T_g)$ it is comparable.
Melt liquidus temperatures are predicted to within \textasciitilde{}60 K, an uncertainty
that approaches those of dedicated polynomial and machine learning
models \citep[e.g.,][]{dreyfus2003}. The melt heat capacities are
predicted within \textasciitilde{}3\%, a precision better than that of 5\% achieved when
using the model of Richet and Bottinga \cite{richet1985} corrected with the
Al\textsubscript{2}O\textsubscript{3}
partial $C_{p}^{liquid}$ value of
Courtial and Richet \cite{courtial1993}, following Giordano and Russell \cite{giordano2017}. Glass
density and refractive index are predicted to within 0.02 g
cm\textsuperscript{-3} and 0.006, respectively. Such values are
comparable to, or better than those for the original i-Melt version for
alkali aluminosilicate compositions \cite{lelosq2021}. For glass
density, the model standard error further compares very well with those
of dedicated parametric \citep[e.g., 0.02 in][]{fluegel2007a} or machine learning \citep[e.g., 0.02 to 0.03 in][]{hu2020} models.
Glass elastic modulus is predicted to within 6 GPa, an accuracy that approaches those achieved by topological models
\cite[e.g.,][]{wilkinson2019} but is higher than that of dedicated machine learning models \citep[e.g., 3 GPa in][]{hu2020}.

\begin{figure}[ht]
    \centering
    \includegraphics{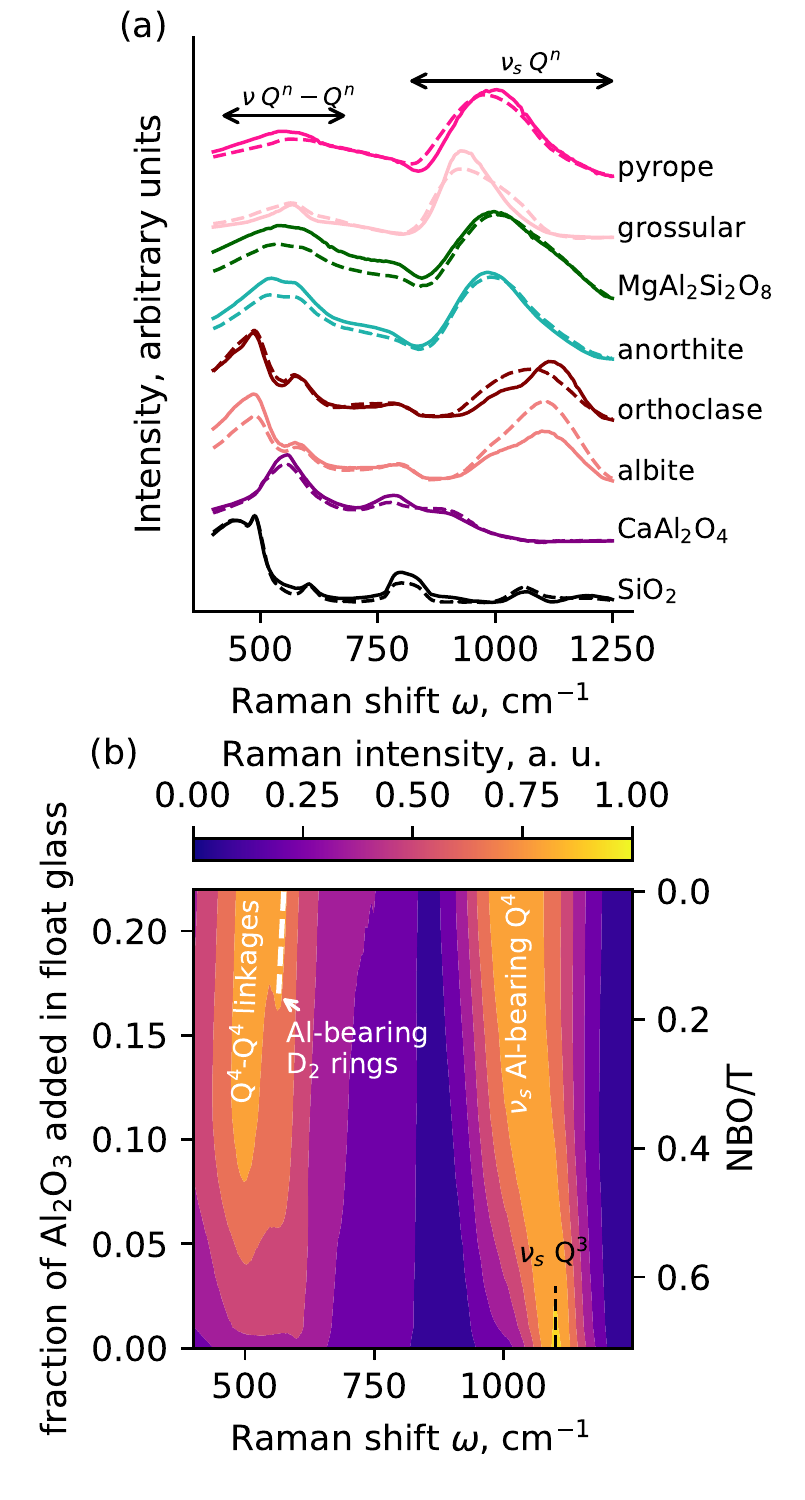}
    \caption{(a) Examples of predicted Raman spectra for specific compositions (dashed curves) represented on top of the measured ones (solid lines).  Ranges of T-O-T (T = Si, Al) intertetrahedral ($Q^n$-$Q^n$) vibrations and Si-O stretching vibrations in $Q^n$ units are indicated at the top. (b) 2D contour plot of the Raman spectral intensity as a function of Raman shift and of the fraction of Al$_2$O$_3$ (mol \%) that was added into a soda-lime silicate float glass composition.}
    \label{fig:raman}
\end{figure}

Global variations in glass Raman signals are well-captured despite
the very small experimental Raman dataset (\textbf{Fig. \ref{fig:fig1}b}): the median absolute
percentage errors (MAPEs) on the training and validation subsets are both equal to \textasciitilde{}16 \%. The MAPE on the testing data subset is of 25 \%. Such errors are comparable to those
affecting the original version of i-Melt. \textbf{Figure \ref{fig:raman}a} shows
examples of Raman spectra predictions for specific compositions, including silica, calcium aluminate,  Ca-Mg silicates and aluminosilicates. The general shape of the spectra
is well reproduced by the model.  This indicates that the model captured well general relationships between glass Raman spectra and their composition.  In detail, there remains visible
deviations between predictions and observations, particularly for small
features in the spectra such as small shoulders and peaks. To get better
Raman spectra predictions, the model could benefit from a broader
training dataset, as this actually is one of the smallest training
dataset. 

While the model shows small deviations from experimental Raman data, it still enables a detailed and systematic exploration of variations in Raman spectra as a function of glass composition. We demonstrate this capability by mapping the changes in Raman intensity upon the addition of Al$_2$O$_3$ into a typical float glass composition (\textbf{Fig. \ref{fig:raman}b}). At null or very low Al$_2$O$_3$ concentrations, a strong Raman signal intensity near 1100 cm$^{-1}$ is observed, which can be assigned to Si-O stretching in $Q^3$ units \citep[][]{mysen1982,mcmillan1984,zotov1999,spiekermann2013,lelosq2022}. This suggests a dominant presence of $Q^3$ units in the soda-lime silicate glass,  This agrees with experimental Raman data \citep{shan2017} and molecular dynamic simulations \cite{cormier2011}, which both indicate a high fraction of $Q^3$ units and lower fractions of $Q^2$ and $Q^4$ units in soda-lime silicate glasses.

Upon the addition of Al$_2$O$_3$ in the float glass composition, the average number of non-bridging oxygens per tetrahedral units (NBO/T) calculated from composition decreases. A significant increase in intensity near 475 cm$^{-1}$ is observed (\textbf{Fig. \ref{fig:raman}b}), indicating the formation of more and more $Q^4$-$Q^4$ bridges as the NBO/T approaches 0 (tectosilicate composition) \cite{mcmillan1984,sharma1985a}. Simultaneously, the Raman signal in the 800-1300 cm$^{-1}$ range undergoes a shift to approximately 1030 cm$^{-1}$ while diminishing in intensity. This particular signal at 1030 cm$^{-1}$ corresponds to the stretching of T-O (T = Si, Al) bonds within $Q^4$ units containing both silicon and aluminum. In pure silica, this vibrational mode of $Q^4$ units is typically observed at around 1200 cm$^{-1}$, but here the presence of aluminum in $Q^4$ units causes a decrease in its frequency \cite{neuville1996,neuville2004}.

The model predictions also accurately reproduce structural details of Al-rich glasses. Above a fraction of approximately 0.18 Al$_2$O$_3$, a signal near 560 cm$^{-1}$ emerges, corresponding to breathing vibrations of three-membered tetrahedral rings ($D_2$ peak) in polymerized aluminosilicate glasses \cite{sharma1985a,kubicki1993,lelosq2013,lelosq2017a}. Typically, this vibrational mode yields a signal near 606 cm$^{-1}$ in silica \cite{galeener1982a,kubicki1993,pasquarello1998,umari2002,umari2003,rahmani2003}. The model's prediction of this signal in the SiO$_2$ spectrum aligns well with experimental observations (\textbf{Fig. \ref{fig:raman}a}), along with the $D_1$ signal near 490 cm$^{-1}$ assigned to breathing vibrations of four-membered rings \cite{sharma1981,galeener1982a,galeener1982, galeener1984,pasquarello1998,umari2002,umari2003,rahmani2003}. The addition of Al into the glass structure leads to a decrease in the frequency of the $D_2$ signal as Al replaces Si in the three-membered rings \cite{sharma1985a,kubicki1993}. This explains the observed $D_2$ signal frequency in the Al-rich soda-lime silicate glass (\textbf{Fig. \ref{fig:raman}b}). 

Overall, this example illustrates that the model Raman predictions document the gradual change in glass structure as glass composition is modified. They also allow for the observation of specific signals, such as those assigned to vibrations in n-membered rings in polymerized glasses. These findings can be combined with property predictions to gain a better understanding of how chemical and structural changes influence the properties of melts and glasses.

\begin{figure}
    \centering
    \includegraphics{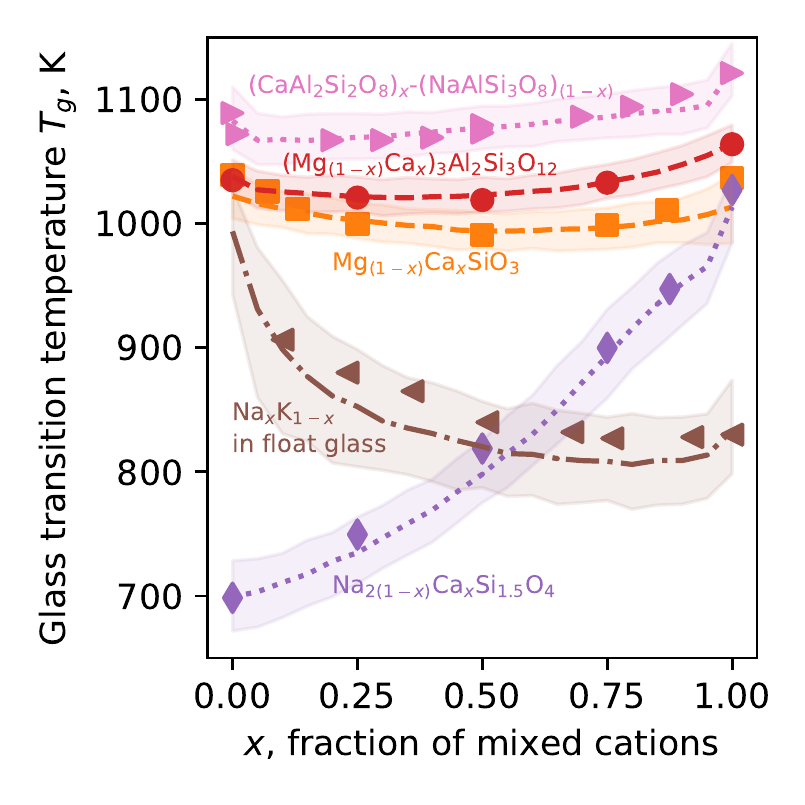}
    \caption{Comparison between data and model predictions of the glass transition temperature $T_g$ of different glass series in which Na-K, Ca-Mg and Na-Ca are mixed, following a mixing fraction $x$. Symbols are data from \cite{hummel1985,neuville1991, neuville2006,kjeldsen2014a}. Lines are model predictions, and shaded areas of the same color represent 95 $\%$ confidence intervals on predictions.}
    \label{fig:mae}
\end{figure}

The versatility of i-Melt in capturing both global and specific variations of properties extends well beyond Raman spectra.  Indeed,  while the different metrics all demonstrate accurate predictions of overall melt and glass properties  (\textbf{Table \ref{tab:tab1}}),  the model also successfully reproduces specific variations in transport and thermodynamic properties, such as the mixed modifier effect (MME). The latter is characterized by extrema in melt/glass properties, such as glass transition temperature or electrical conductivity, when different metal cations are mixed \citep[e.g., see the reviews of][]{isard1969,day1976,greaves2007}. In details, the MME exhibits some subtleties and is not necessarily easy to reproduce accross a broad range of compositions. For instance, mixing Na and K in silicate compositions results in a pronounced MME on $T_g$ and viscosity \cite{poole1949c,richet1984c},  whereas mixing these cations in aluminum-rich melt series does not induce a significant MME \cite{lelosq2013,lelosq2017a}. The MME on viscosity is not observed when mixing Na and Ca in silicate melts \cite{neuville2006a},  but it is observed in feldspatic albite-anorthite aluminosilicate melts \cite{hummel1985}. The earlier version of i-Melt already successfully captured such subtleties for Na-K aluminosilicate compositions \cite{lelosq2021}, confirming that the introduction of aluminum leads to a change in the Na-K mixing behavior. In \textbf{figure \ref{fig:mae}}, we present the glass transition temperature ($T_g$) for different data series involving mixing of Ca-Mg, Ca-Na, and Na-K metal cations in silicate and more complex aluminosilicate compositions, including float glass. The model predictions closely match the data, all falling within the 95\% predictive confidence intervals. The model accurately predicts the mixing effect of Ca and Mg on $T_g$ in silicate and aluminosilicate compositions, as well as the Na-K mixing effect in float glass. Additionally, the variations of $T_g$ upon mixing Na and Ca in silicate glasses or along the anorthite-albite binary are also well reproduced. i-Melt effectively captures the detailed variations of melt properties resulting from metal cation mixing. These capabilities prove valuable in exploring how phenomena like the MME systematically change with variations in melt/glass composition. The ability of i-Melt to successfully predict Raman spectra further provides structural insights for such analyses.

\subsection{Model internal consistency}

Among predicted parameters, i-Melt returns the melt
fragility \emph{m}, which corresponds to the derivative of melt
viscosity against temperature at $T_g$. Melt
fragility is related to structural and thermodynamic melt
properties, and allows distinguishing ``strong'' melts from ``fragile''
ones that show a strongly non-Arrhenian dependence of their viscosity
against temperature \cite{angell1991}. In particular, melt fragility is
expected to correlate with melt thermodynamic properties \cite{toplis1997a}. 
This can be explored within the framework of the Adam-Gibbs
theory \cite{adam1965}, which relates melt dynamic and
thermodynamic properties through the equation \cite{richet1984c}:

\begin{equation}
    \log_{10}\eta = A_{e} + \frac{B_{e}}{T [S^{conf}(T_{g}) + \int_{T_g}^{T} \frac{C_{p}^{conf}}{T}\,dT ] }
    \label{eq:eq1}
\end{equation}

with \emph{T (K)} the temperature, \emph{A\textsubscript{e}} (Pa·s) a
high-temperature limit, \emph{B\textsubscript{e}} (J
mol\textsuperscript{-1}) a term proportional to the energy barriers
opposed to molecular rearrangements, and
$S^{conf}$ (J mol\textsuperscript{-1}
K\textsuperscript{-1})
and $C_p^{conf}$ (J
mol\textsuperscript{-1} K\textsuperscript{-1}) the melt configurational entropy and heat
capacity, respectively.
$C_p^{conf}(T)$ is equal to \cite{richet1984c}:

\begin{equation}
    C_{p}^{conf}(T) = C_{p}^{liquid}(T) - C_{p}^{glass}(T_{g})
    \label{eq:eq2}
\end{equation}

with $C_{p}^{glass}(T_{g})$ the glass heat capacity at T\textsubscript{g}
and $C_{p}^{liquid}(T)$ the liquid heat capacity at T. In the model,
$C_{p}^{glass}(T_{g})$ is well predicted via the Dulong and Petit limit (see
discussion in \cite{richet1987}). $C_{p}^{liquid}(T)$ is predicted from the
ponderated sum of neural network predicted partial molar heat capacities
of the oxide components, and neural network predicted temperature terms
for Al\textsubscript{2}O\textsubscript{3} and K\textsubscript{2}O (see
for a discussion regarding those terms \cite{richet1985,courtial1993}).

Following the relationship between melt viscosity and thermodynamic
properties (eq. \ref{eq:eq1}), melt fragility should be proportional to the ratio
of the configurational heat capacity at $T_g$ over
the configurational entropy at \emph{T\textsubscript{g},
C\textsubscript{p}\textsuperscript{conf}(T\textsubscript{g})/S\textsuperscript{conf}(T\textsubscript{g})}, as \cite{toplis1997a}:

\begin{equation}
    m \propto 1 + \frac{C_{p}^{conf}(T_{g})}{S^{conf}(T_{g})}
    \label{eq:eq3}
\end{equation}

\begin{figure}
    \centering
    \includegraphics{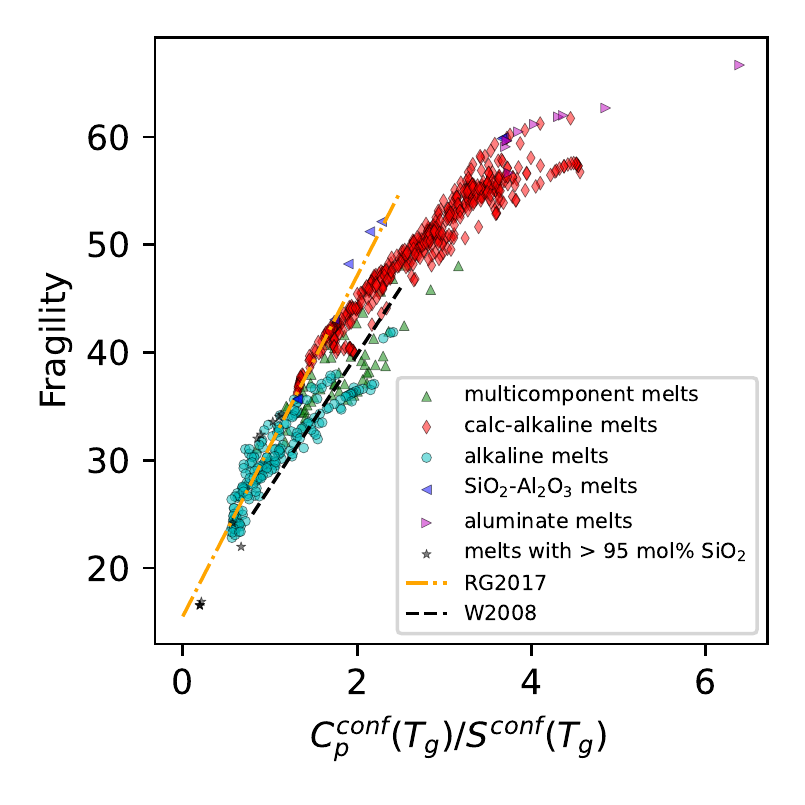}
    \caption{Glass fragility versus melt $C_{p}^{conf}(T_{g})/S^{conf}(T_{g})$ ratio. Symbols are predictions of the machine learning framework on the different subsets of the \emph{D\textsubscript{viscosity}} dataset. The dashed line is the relationship observed by Webb (\cite{webb2008a}, abbreviated W2008 in the figure legend) using experimental heat capacity data, and the dashed-dotted line is that observed by Russell and Giordano (\cite{russell2017}, abbreviated RG2017 in the figure legend). }
    \label{fig:consistency}
\end{figure}

The correlation expected from eq. \ref{eq:eq3} has been experimentally observed
for melts presenting relatively restrained compositional ranges \cite{webb2008a,russell2017}. 
Therefore, if i-Melt is internally
consistent, i.e. if it predicts physically realistic melt fragilities,
heat capacities and configurational entropies, we should also observe
the correlation expected from eq. \ref{eq:eq3}. This is clearly the case, as
observed in \textbf{Figure \ref{fig:consistency}}. We observe a general trend located in
between those reported in \cite{webb2008a} and \cite{russell2017},
the first study concerning sodium and calcium aluminosilicate melts and
the second one iron-bearing aluminosilicate melts of geological
compositions.

In the past version of i-Melt, focused on predictions in the
Na\textsubscript{2}O-K\textsubscript{2}O-Al\textsubscript{2}O\textsubscript{3}-SiO\textsubscript{2} system,
the correlation expected from eq. \ref{eq:eq3} was observed but with a significant
scatter (see Figure 6 in \cite{lelosq2021}). Here, the scatter is
much more limited, despite the fact that predictions cover a broader
compositional space. A systematic, albeit not fully linear trend between
melt fragility
and \emph{C\textsubscript{p}\textsuperscript{conf}(T\textsubscript{g})/S\textsuperscript{conf}(T\textsubscript{g})} is
observed (\textbf{Fig. \ref{fig:consistency}}). The improvement in internal consistency is
actually due to predicting \(C_{p}^{liquid}(T)\) through the artificial
neural network. Indeed, we tried modeling \(C_{p}^{liquid}\) using
partial molar liquid heat capacity values for the different oxide
components from several studies \citep[e.g.,][]{richet1985,courtial1993,tangeman1998,webb2008a}.
However, the present predictive range of the model covers a very wide
compositional range (\textbf{Fig. \ref{fig:fig1}}), including Na-rich and Mg-rich
compositions for which non-linear dependence of partial molar \(C_{p}^{liquid}\)
has been reported \cite{courtial1993,tangeman1998}.
Artificial neural network predictions of partial molar
\(C_{p}^{liquid}\) for the different oxide components, and their
associated temperature dependence for
Al\textsubscript{2}O\textsubscript{3} and K\textsubscript{2}O, allowed
largely increasing the consistency of the model, while yielding good
predictions of \(C_{p}^{liquid}\left(T\right)\) (\textbf{Figs. \ref{fig:errors}e, \ref{fig:consistency}} ).

In \textbf{figure \ref{fig:consistency}}, we observe a clear compositional
mapping between melt fragility
and $C_p^{conf}(T_g)/S^{conf}(T_{g})$. Si-rich and alkali-bearing compositions systematically present low
fragilities and $C_p^{conf}(T_g)/S^{conf}(T_{g})$, while 
calc-alkaline compositions present higher ones. Ca aluminate
compositions present the highest fragilities
and $C_p^{conf}(T_g)/S^{conf}(T_{g})$ (magenta
symbols in \textbf{Fig. \ref{fig:consistency}}). The general trend between
melt fragility and $C_p^{conf}(T_g)/S^{conf}(T_{g})$ is
almost linear, but deviates from linearity at high fragilities
and $C_p^{conf}(T_g)/S^{conf}(T_{g})$ ratios.

\subsection{Uncertainty estimations and model extrapolation}

Obtaining uncertainty estimations from machine learning models can be
challenging. While several methods exist to address this issue, there is
no one-size-fits-all approach for obtaining uncertainty estimations. In
the case of artificial neural networks, there are at least three established
methods for obtaining uncertainty estimations on model predictions.
These include using Bayesian neural networks \cite{kononenko1989, izmailov2021}, MC dropout \cite{gal2016}
or conformal predictions \citep[see][and references therein]{angelopoulos2021}.

Here, we performed a systematic analysis of the confidence intervals
provided using the MC Dropout method, which relies on the use of
dropout. The latter consists in randomly turning off activation units of
the artificial neural network during training, at each iteration with a
probability \emph{p}. This regularization method is recognized to help
generalization in deep learning models. At inference, it is also
possible to use dropout to generate multiple, slightly different
predictions for a given input. The obtained sample of predictions can
then be used to produce an estimate of uncertainties \cite{gal2016}.

To assess the reliability of the uncertainties estimated using MC
Dropout, we computed the 2.5th and 97.5th percentiles of the predictive
distributions for the different test data subsets and compared them to
the corresponding observed values (\textbf{Table \ref{tab:tab2}}). Ideally, the
calculated confidence intervals should encompass 95 \% of the
observations in the datasets. For all properties but Raman spectra, Elastic modulus and liquid heat capacity,  the 95 \% confidence intervals calculated with MC Dropout encompass between 84 and 90 \% of the test data. The proportion of data included in the 95\% confidence intervals is lower for Raman spectra,  Elastic modulus and liquid heat capacity.
This indicates that calculated confidence intervals using MC Dropout are generally too narrow.  MC Dropout thus provides first-order estimates of confidence intervals, but those may not perfectly cover the
true ones, particularly for Raman spectra,  Elastic modulus and liquid heat capacity.

\begin{table}[]
    \centering
    \begin{tabular}{c|c|c}
        \toprule
         Dataset & MC dropout & Conformal c.i.\\
        \midrule
        Viscosity            & 90 \% & \textbf{94 \%}\\
        Density              & 87 \% & \textbf{87 \%}\\
        Raman spectra        & 70 \% & \textbf{97 \%}\\
        Refractive index     & 90 \% & \textbf{97 \%}\\
        CTE                  & 84 \% & \textbf{97 \%}\\
        Elastic Modulus      & 76 \% & \textbf{94 \%}\\
        Abbe number          & \textbf{81 \%} & 78 \%\\
        Liquidus temperature & 85 \% & \textbf{96 \%}\\
        Heat capacity        & 57 \% & \textbf{86 \%}\\
        \bottomrule
    \end{tabular}
    \caption{Percentage of test data within the 95\%
confidence intervals (c.i.) calculated using MC Dropout (MC dropout
column), or after MC Dropout c.i. calibration using conformal prediction
(Conformal c.i. column). Numbers closest to 95\% are good.  }
    \label{tab:tab2}
\end{table}

Fortunately, for observed properties, it is possible to effectively
scale MC Dropout confidence intervals using conformal predictions,
following the Time-Test Dropout method proposed by Cortes-Ciriano and
Bender \cite{cortes-ciriano2019}. Conformal prediction uses the observed
deviations between predictions and observations from a calibration dataset
to calculate reliable confidence intervals for new, unseen values.
Here, we use the validation data subsets as calibration datasets. We
estimate MC dropout confidence intervals on those calibration datasets,
and forward them to a conformal regressor model. The latter will make
use of the calibration data to effectively scale the confidence
intervals provided by the MC dropout method.

Using this method, 95 \% of the test data, or more in some cases, now fall
within the 95\% scaled confidence intervals (\textbf{Table \ref{tab:tab2}}), with the
exception of the Abbe number for which MC Dropout uncertainties appear
to be best. Therefore, for most properties, the
model conformal confidence intervals are reliable, even slightly conservative (i.e. the
95\% confidence intervals actually encompass more than 95\% of the
data). Therefore, for observed properties, scaling the MC Dropout
confidence intervals via conformal prediction allows an effective
re-scaling of the confidence intervals. Unfortunately, such scaling is not possible for
latent properties, for which MC Dropout still allows obtaining a reasonable
approximation of their confidence intervals.

\begin{figure}
    \centering
    \includegraphics[width=\textwidth]{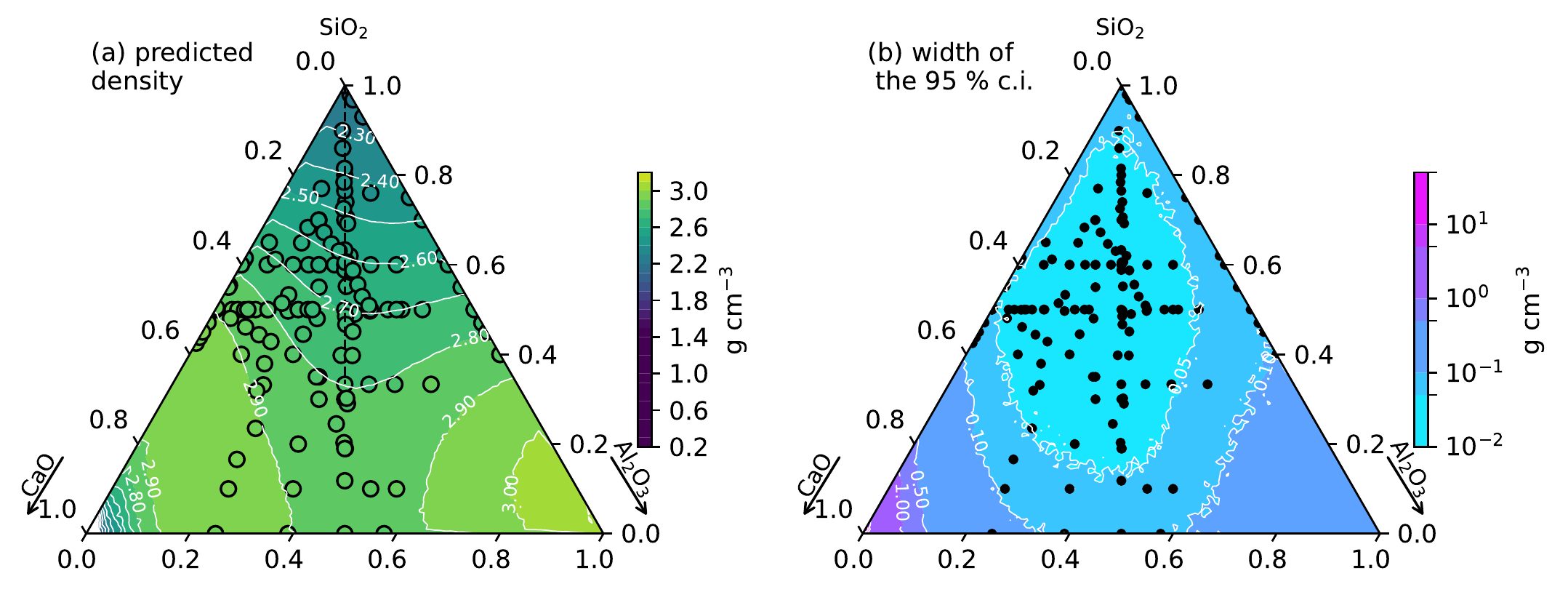}
    \caption{Ternary diagrams of predicted density of CaO-Al$_2$O$_3$-SiO$_2$ glasses (a) and of the width of the associated 95\% conformal confidence intervals (b).  Circles in (a) and (b) are experimental data points from the i-Melt database.  In (a), the colors of those circles scales with the measured density.  In general, no large color difference between the data points and the background,  which represents model predictions, is observed because of the high model precision. The color bar in (b) follows a log scale to better highlight the regions of high and low values.}
    \label{fig:uncertainties}
\end{figure}

The ability of predicting confidence intervals is important, as of course
it allows estimating model uncertainties, but maybe more importantly, it
allows detecting model extrapolation. To illustrate this, we show
in \textbf{Figure \ref{fig:uncertainties}} model predictions of glass density
 in the CaO-Al\textsubscript{2}O\textsubscript{3}-SiO\textsubscript{2} ternary system., together with existing data. We observe that the 95 \% predictive conformal confidence intervals are lower than 0.05 g cm$^{-3}$ when the density of data is the highest, near the center of the ternary diagram (\textbf{Fig. \ref{fig:uncertainties}}b). When the data become scarce, the 95\% confidence intervals increase to values in between 0.05 and 0.1 g cm$^{-3}$,  while in regions where no data are present, i.e. at very high CaO or Al$_2$O$_3$ concentrations, the model 95 \% predictive conformal confidence intervals quickly become high, with values that can reach more than 1 g cm$^{-3}$ (\textbf{Fig. \ref{fig:uncertainties}}b). This indicates that the model is
in the extrapolative regime, and that predictions are subject to
(very) significant uncertainties.  Overall, the present results
demonstrate that, in addition to delivering trustworthy interpolative
uncertainties, the methodology incorporated into the updated version of
i-Melt allows the identification of poor extrapolation performance by a
considerable increase in predicted uncertainties.  

Finally,  the utilization of model predictive confidence intervals can offer valuable insights for designing future experiments, employing a similar approach to Bayesian experimental design \citep[e.g.,][]{chaloner1995}. Specifically,  high model confidence intervals often indicate regions with limited data (\textbf{Fig. \ref{fig:uncertainties}}b). Therefore,  compositional areas characterized by significant uncertainties in predictions are identified as prime candidates for further experimentation, as they hold the potential to refine and enhance the model's accuracy. Consequently, these areas represent promising avenues for experimental investments. Models such as i-Melt, therefore, serve as valuable tools in maximizing the expected utility of the data obtained from future experiments.

\section{Discussion}

The inclusion of MgO and CaO in the i-Melt model has been accomplished
without compromising its predictive accuracy. In fact, as shown
in \textbf{Table \ref{tab:tab1}}, the error metrics for the extended model are
comparable to, or even better than those of the original version, which
focused only on the
Na\textsubscript{2}O-K\textsubscript{2}O-Al\textsubscript{2}O\textsubscript{3}-SiO\textsubscript{2} system
and predicted less melt/glass properties. This is attributed to several
factors, including a significantly enlarged database, the incorporation
of new chemical descriptors as inputs to the artificial neural network,
and the use of GELU activation units with a larger dropout rate,
enabling moderately deep artificial neural networks to generalize well
and produce accurate predictions. Consequently, the updated model
enables systematic predictions of properties for melts and glasses in
the
Na\textsubscript{2}O-K\textsubscript{2}O-MgO-CaO-Al\textsubscript{2}O\textsubscript{3}-SiO\textsubscript{2} system.
It covers a broad compositional range that encompasses the entire
glass-forming domain of this system and its related sub-systems
(\textbf{Fig. \ref{fig:fig1}}). Furthermore, in addition to high predictive
performances (\textbf{Figs. \ref{fig:errors}, \ref{fig:raman}}), the model provides reliable
uncertainty estimates using MC Dropout and conformal predictions (\textbf{Table \ref{tab:tab2}, Fig. \ref{fig:uncertainties}}).

Future extension of the model may include the addition of new properties
and oxide elements. However, it should be noted that the addition of a
new oxide element may require significantly more compositions to be
added to the database to maintain the achieved level of precision. At
the moment, it is difficult to estimate how much data needs to be added
to maintain the model precision when adding new elements. Despite this,
the fact that the new version of i-Melt is as precise as, or better than
the original one that focused on a quaternary system is very
encouraging. This suggests that this problem may not be critical. One
point will require a particular focus: the addition of Raman spectra.
This dataset is currently the smallest, and higher predictive precision
may be achieved by adding more Raman spectra. Data on heterogeneous
samples, such as Raman maps along diffusion profiles, could largely
benefit model predictions, and are very welcomed.

With the new version of i-Melt, it is possible to predict many different
latent and observed thermodynamic, physical and structural properties of
glasses and melts. This opens up a range of possibilities for studying
alkali and calc-alkaline aluminosilicate melts and glasses, such as
exploring correlations between structure and properties \citep[see e.g. \textbf{Figs. \ref{fig:fig_visco}, \ref{fig:raman}} and other examples in][]{lelosq2021}, investigating
properties in multicomponent systems (e.g., \textbf{Fig. \ref{fig:uncertainties}}), and studying cation mixing effects (e.g., \textbf{Fig. \ref{fig:mae}}).  
These possibilities were highlighted in \cite{lelosq2021}, 
and examples are provided in the i-Melt repository online. Here, for the sack of concision, we
focus on the noteworthy ability of the new version to predict
partial molar volumes and heat capacities of oxide components, which has
important implications for understanding the thermodynamics and
transport properties of melts and glasses.

Predicted partial molar volumes of oxide components fall close to those
reported in previous publications (\textbf{Table \ref{tab:tab3}}), but the model
predicts that they depend on composition. A similar comment can be made
for partial molar $C_{p}^{liquid}$.
Such a compositional dependence is not surprising, because it has been
reported in several publications for density \cite[e.g., ][and references therein]{neuville2022a} and for heat capacity \cite{courtial1993, tangeman1998}.

\begin{table}[]
    \centering
    \begin{tabularx}{\textwidth}{c|c|c|c|c|c|c}
         \toprule
         & \multicolumn{4}{c|}{Partial molar volume, cm$^{3}$ mol$^{-1}$} & \multicolumn{2}{c}{$C_p^{liquid}$, J mol$^{-1}$ K$^{-1}$}\\
         Oxide & 2.5\textsuperscript{th} & 50\textsuperscript{th} & 97.5\textsuperscript{th} & Literature values & Average model values & Literature values\\
         \midrule
         SiO$_2$ & 24.1 & 25.7 & 27.1 & 26.0-27.5 & 80 $\pm$ 2 & 81.37\\
         Al$_2$O$_3$ & 34.5 & 37.8 & 38.4 & 37.0-39.0 & $118\pm5 + 0.028\pm0.006\times T$ & $130.2 + 0.0357\times T$\\
         Na$_2$O & 21.3 & 22.9 & 29.1 & 25.0-29.0 & 99 $\pm$ 7 & 100.6\\
         K$_2$O & 34.7 & 36.9 & 50.1 & 40.0-46.0 & $71\pm14 + 0.017\pm0.005\times T$ & $50.13 + 0.01578\times T$\\
         MgO & 8.1 & 13.1 & 16.5 & 11.0-13.0 & 81 $\pm$ 8 & 85.78\\
         CaO & 14.0 & 14.9 & 18.7 & 14.0-18.0 & 92 $\pm$ 11 & 86.05\\
         \bottomrule
    \end{tabularx}
    \caption{2.5, 50 and 97.5 quantiles of partial molar volumes $V_m$ and average liquid heat capacities $C_p^{liquid}$ of oxide components calculated using the density and heat capacity datasets. Here, important error bars indicate compositional dependence of those values, and do not reflect model uncertainties. Ranges of reported values are from \cite{bottinga1983, lange1987, liu2006, neuville2022a} for partial molar volumes and \cite{richet1985, courtial1993} for partial liquid heat capacities.}
    \label{tab:tab3}
\end{table}

The model allows understanding of the potential origin of such
non-linear dependence on composition. For
partial $V_{m}$, an explanation may come from changes
in the average oxygen coordination number (CN) of the cations. For
instance, increasing
Al\textsubscript{2}O\textsubscript{3}/M\textsuperscript{x+}\textsubscript{2/x}O
in aluminosilicate glasses induces a change in the role of the metal
cation M\textsuperscript{x+}\textsubscript{2/x}, from network modifier
(breaking Si-O-Si bonds) to charge compensator of
Al\textsuperscript{3+} \citep[][]{mysen2003a, lelosq2014, bechgaard2017}. In sodium aluminosilicate glasses with 75 mol\%
SiO\textsubscript{2}, this correlates with an increase in
Na\textsuperscript{+} CN as the
Al\textsubscript{2}O\textsubscript{3}/(Na\textsubscript{2}O+Al\textsubscript{2}O\textsubscript{3})
ratio increases, as documented by \textsuperscript{23}Na Nuclear
Magnetic Resonance (NMR) spectroscopy data \cite{lelosq2014}. A
similar conclusion was drawn from calorimetry data on alkali
aluminosilicate compositions by \cite{richet1993b}, who proposed an
increase in Na\textsuperscript{+} CN from 6 to 9 as Al is introduced
into the glass network. Therefore, for metal cations such as
Na\textsuperscript{+}, we may expect a distribution of
partial $V_{m}$
and $C_{p}^{liquid}$ values as a
function of glass composition, due to changes in the local environment
of the cations.

\begin{figure}
    \centering
    \includegraphics[width=\textwidth]{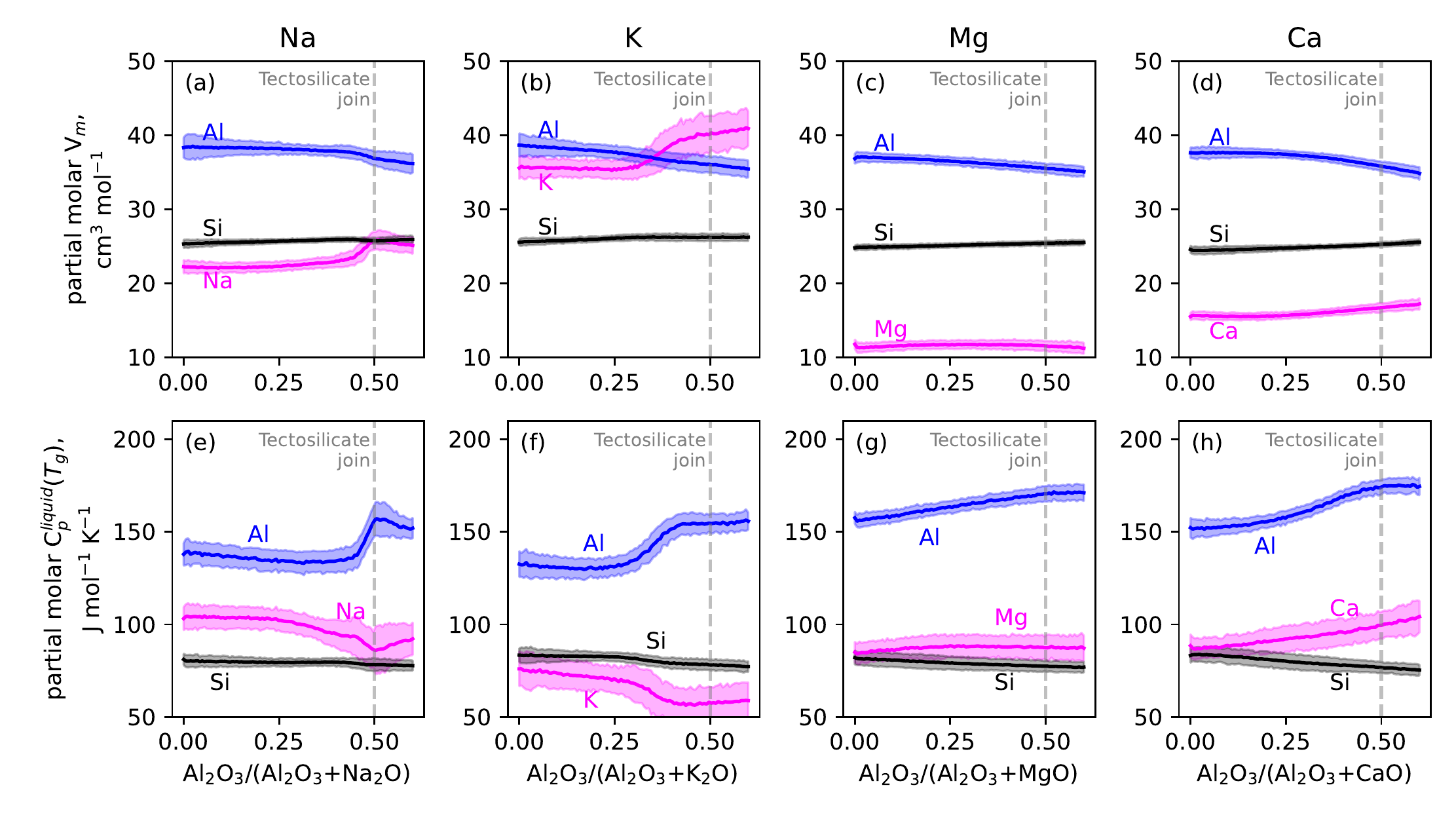}
    \caption{Predicted partial molar volumes in glasses and
partial molar contributions to liquid heat capacity of
SiO\textsubscript{2}, Al\textsubscript{2}O\textsubscript{3} and metal
cations M\textsubscript{2/x}\textsuperscript{x+}O as a function of the
Al\textsubscript{2}O\textsubscript{3}/(Al\textsubscript{2}O\textsubscript{3}+
M\textsubscript{2/x}\textsuperscript{x+}O) ratio in
Na\textsubscript{2}O-Al\textsubscript{2}O\textsubscript{3}-SiO\textsubscript{2} (panels
a and e),
K\textsubscript{2}O-Al\textsubscript{2}O\textsubscript{3}-SiO\textsubscript{2} (panels
b and f), MgO-Al\textsubscript{2}O\textsubscript{3}-SiO\textsubscript{2}
 (panels c and g) and
CaO-Al\textsubscript{2}O\textsubscript{3}-SiO\textsubscript{2} (panels d
and h) compositions. Mole fraction of SiO\textsubscript{2} is fixed at
0.5. The shaded areas represent 95 \% confidence intervals calculated
with MC Dropout, and the solid lines the median values of model
predictions.}
    \label{fig:vm}
\end{figure}

To test this hypothesis, we explore how the partial
molar $V_{m}$
and $C_{p}^{liquid}$ of
M\textsuperscript{x+}\textsubscript{2/x} O,
Al\textsubscript{2}O\textsubscript{3} and SiO\textsubscript{2} vary with
the ratio \emph{X\textsubscript{Al}} =
Al\textsubscript{2}O\textsubscript{3}/(M\textsuperscript{x+}\textsubscript{2/x}O+Al\textsubscript{2}O\textsubscript{3})
in glasses containing 50 mol\% SiO\textsubscript{2} (\textbf{Fig. \ref{fig:vm}}).
Upon increasing \emph{X\textsubscript{Al}} , the partial molar
$V_{m}$ of SiO\textsubscript{2} barely varies
(\textbf{Fig. \ref{fig:vm}a,b,c,d}). That of Al\textsubscript{2}O\textsubscript{3}
decreases from \textasciitilde{}39 cm\textsuperscript{3}
mol\textsuperscript{-1} down to \textasciitilde{}35
cm\textsuperscript{3} mol\textsuperscript{-1}
when \emph{X\textsubscript{Al}} becomes larger than 0.3.
Furthermore, $V_{m}$
Al\textsubscript{2}O\textsubscript{3} appears to be slightly lower in
calc-alkaline systems (\textbf{Fig. \ref{fig:vm}c,d}) than in alkaline ones
(\textbf{Fig. \ref{fig:vm}a,b}) when \emph{X\textsubscript{Al}}\textless{}
\textasciitilde{}0.3. The partial $V_{m}$ of
Na\textsubscript{2}O and K\textsubscript{2}O both significantly increase
by \textasciitilde{}3 and \textasciitilde{}5 cm\textsuperscript{3}
mol\textsuperscript{-1} (\textbf{Fig. \ref{fig:vm}a,b} ), respectively, when
\emph{X\textsubscript{Al}} becomes higher than 0.3. The partial
\emph{V\textsubscript{m}} of MgO does not vary significantly with
\emph{X\textsubscript{Al}} (\textbf{Fig. \ref{fig:vm}c}) while that of CaO
slightly increases by \textasciitilde{}2
cm\textsuperscript{3}mol\textsuperscript{-1} (\textbf{Fig. \ref{fig:vm}d}).

Similar observations can be made for
partial $C_{p}^{liquid}$ values
(\textbf{Fig. \ref{fig:vm}}). Whereas those of SiO\textsubscript{2}
(\textbf{Fig. \ref{fig:vm}e,f,g,h}) and MgO (\textbf{Fig. \ref{fig:vm}g}) barely vary, that
of CaO slightly increases by \textasciitilde{}15 J
mol\textsuperscript{-1} K\textsuperscript{-1} as
\emph{X\textsubscript{Al}} increases (\textbf{Fig. \ref{fig:vm}h}). The
partial $C_{p}^{liquid}$ values of
Na\textsubscript{2}O and K\textsubscript{2}O show distinct decreases of
\textasciitilde{}11 and \textasciitilde{}17 J
mol\textsuperscript{-1}K\textsuperscript{-1}, respectively, as
\emph{X\textsubscript{Al}} increases (\textbf{Fig. \ref{fig:vm}e,f}). Turning to
the partial $C_{p}^{liquid}$ of
Al\textsubscript{2}O\textsubscript{3}, it increases by
20 to 25 J mol\textsuperscript{-1} K\textsuperscript{-1}
as \emph{X\textsubscript{Al}} increases in alkali melts, with a marked
step at \emph{X\textsubscript{Al}} \textasciitilde{} 0.4 (\textbf{Fig.
\ref{fig:vm}e,f}). In calcalkaline
melts, $C_{p}^{liquid}$ Al\textsubscript{2}O\textsubscript{3}
is significantly higher, and nearly linearly increases with
\emph{X\textsubscript{Al}} from \textasciitilde{}151 J
mol\textsuperscript{-1} K\textsuperscript{-1} to \textasciitilde{}172 J
mol\textsuperscript{-1} K\textsuperscript{-1}.

Based on previous works \cite{richet1993b, cormier2004, lelosq2014}, we hypothesize that the observed
$V_{m}$ changes in alkali systems occur as
K\textsuperscript{+} and Na\textsuperscript{+} forms compensating
complexes with AlO\textsubscript{4}\textsuperscript{5-} tetrahedral
units: this results in variations in the CNs of the alkali metal cations
that may induce variations in their partial molar volume. For
alkaline-earth metal cations, we expect limited variations in their CNs \cite{cormier2004,gambuzzi2015,deng2022},
explaining the relatively small variations in their partial
$V_{m}$. In parallel, the change, or lack of change
in the local environments of alkali and alkaline-earth cations also
seems related to the variations, or lack of variations of their
partial $C_{p}^{liquid}$ values
with \emph{X\textsubscript{Al}}.

\begin{figure}
    \centering
    \includegraphics{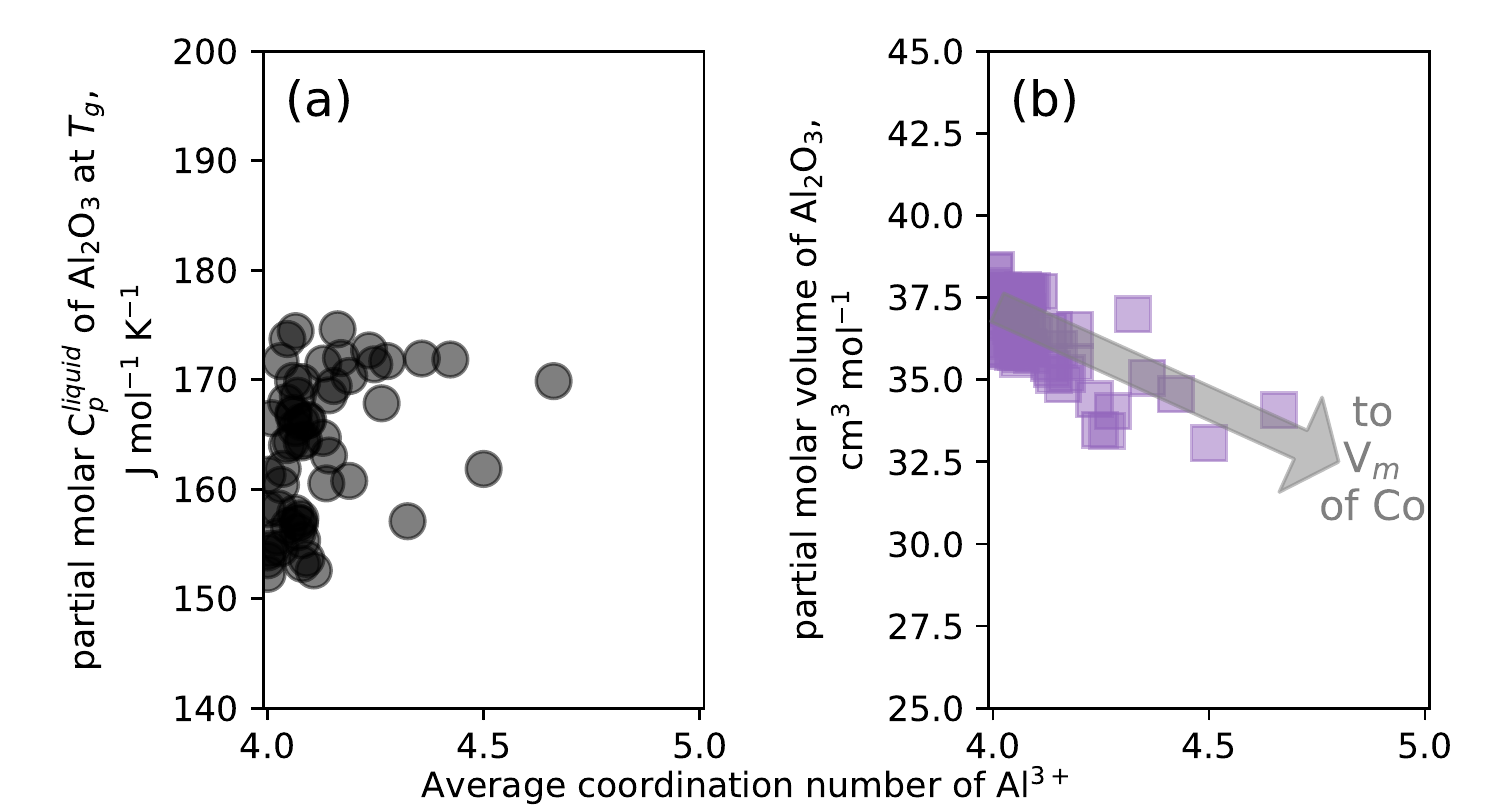}
    \caption{Partial molar contribution of
Al\textsubscript{2}O\textsubscript{3} to the liquid heat capacity at the
glass transition temperature $T_{g}$ (a) and to the
glass molar volume (b) as a function of the mean coordination number of
Al\textsuperscript{3+}.}
    \label{fig:al_vm_cp}
\end{figure}

The case of Al\textsubscript{2}O\textsubscript{3} is more complex.
First, there is a large dependence of partial
Al\textsubscript{2}O\textsubscript{3} $C_{p}^{liquid}$
to temperature: the higher glass transition temperatures of Al-rich and
alkaline-earth bearing melts naturally incur higher values of
Al\textsubscript{2}O\textsubscript{3} $C_{p}^{liquid}$. 
Besides, Al in CN 5 and 6 is detected in aluminium-rich glasses \cite{mcmillan1992,stebbins2000,stebbins2008, toplis2000, neuville2008b, thompson2011, thompson2012, thompson2013,lelosq2014,park2018}, and may further
induce variations in its partial contributions to
$V_{m}$ and $C_{p}^{liquid}$.  
Indeed, such an effect was observed for instance for the heat capacity of calcium aluminosilicate glasses, in which the partial contributions of Al in CN 4, 5 and 6 to the $C_{p}^{glass}$ were estimated to be of 80.3, 79.9 and 70.0 J mol$^{-1}$ K$^{-1}$, at 300 K respectively \cite{richet2009}. To test this hypothesis, we used the compilation of the fractions of Al in CN
4,  CN 5 and CN 6 in magnesium and calcium aluminosilicate glasses made
by \cite{neuville2022a}, and we report the average coordination number of Al against Al\textsubscript{2}O\textsubscript{3} partial
molar $V_{m}$
and $C_{p}^{liquid}$
at $T_{g}$ (\textbf{Fig. \ref{fig:al_vm_cp}}). While no systematic
relationship is observed between partial
Al\textsubscript{2}O\textsubscript{3} $C_{p}^{liquid}$
and the fraction of \textsuperscript{{[}5{]}}Al and
\textsuperscript{{[}6{]}}Al, a systematic trend is observed for
Al\textsubscript{2}O\textsubscript{3} partial molar volume. In general,
Al\textsubscript{2}O\textsubscript{3} $V_{m}$ seems to
tend toward the value of 25.575 cm\textsuperscript{3}
mol\textsuperscript{-1} for corundum, where Al is in CN 6 (\textbf{Fig.
\ref{fig:al_vm_cp}b}). This demonstrates that changes in CN of network-former cations also explain the variations in their partial molar $V_{m}$. On the contrary,  changes in polymerization resulting from variations in \emph{X\textsubscript{Al}} do not seem to induce important variations in the SiO\textsubscript{2} partial molar $V_{m}$ (\textbf{Fig. \ref{fig:vm}}). Indeed, as \emph{X\textsubscript{Al}} increases,  a marked change in the distribution of Si in different $Q^n$ tetrahedral units is expected \cite[e.g.,][]{mysen2003a,lelosq2014},  as recorded for instance in Raman spectra of the float glass in \textbf{figure \ref{fig:raman}b}. Despite this, no large variations in SiO\textsubscript{2} partial molar $V_{m}$ is observed (\textbf{Fig. \ref{fig:vm}a,b,c,d}). This may find an explanation in the fact that the difference in length between Si-BO and Si-NBO bonds is relatively small ($\leq$ \textasciitilde{} 2 \%, e.g. see \cite{ispas2010}), and thus can only result in limited variations in SiO\textsubscript{2} $V_{m}$.

The above discussion indicates that changes in the partial
molar $V_{m}$ of oxide components in glasses occur as
the CNs of cations evolve with glass composition. This process also
should be particularly important in melts, in which changes in the
coordination numbers of cations are occurring not only as a function of
composition, but also as a function of temperature \cite{allwardt2005a, kanehashi2007, neuville2008a, lelosq2014} and pressure \cite{yarger1995,lee2004,lee2004a,allwardt2005, sanloup2013, sanloup2013a, drewitt2015, sanloup2016, lee2020a}. Therefore,
systematic and precise predictions of melt molar volume and density, a
key area to solve questions related to the presence and behavior of deep
silicate melts in the inner Earth near the mantle transition zone or the
mantle-core boundary \citep[e.g.][]{sanloup2016}, requires further
knowledge regarding the links between cationic environment and molar
volumes. For a given cation, combining i-Melt predictions of cationic
$V_{m}$ with experimental data regarding the cationic
environment may result in producing interesting data to better constrain
the links between melt/glass composition, cation partial molar volumes,
local atomic environments and glass/melt density.

\section{Conclusion}

The new implementation of i-Melt, a greybox model combining artificial
neural networks with physical equations, allows systematic and precise
predictions of the properties of alkali and alkaline-earth
aluminosilicate melts and glasses, including configurational entropy,
liquid heat capacity and partial molar contributions from the different
oxide components, configurational heat capacity, glass transition
temperature, fragility, viscosity, density and partial molar volumes of
oxide components, optical refractive index and Raman spectra. Reliable
uncertainty estimates are now provided using MC dropout and conformal
prediction.

The new abilities of the model, i.e. the prediction of partial oxide
values for glass molar volumes and liquid heat capacities, allow
understanding the links between the role and environment of cations in
glasses and melts, and variations in melt/glass properties. For
instance, at fixed SiO\textsubscript{2} but varying
Al\textsubscript{2}O\textsubscript{3}/M\textsubscript{2/x}\textsuperscript{x+}O,
M\textsubscript{2/x}\textsuperscript{x+}O and
Al\textsubscript{2}O\textsubscript{3} partial molar volumes and liquid
heat capacities at $T_g$ can vary as a function of the
role of metal cation M in the glass network (i.e., if acting as a
network modifier or charge compensator), as well as a function of the
fractions of Al in 4, 5 and 6 fold coordination. Such a demonstration is
only a glimpse into the possibilities offered by machine-learning
powered models such as i-Melt. Their development can help better
understand the properties and structure of melts under various
conditions pertinent for geologic and industrial problems, design new experiments,  and new glass products The open source and free nature of i-Melt implies that
future developments also can greatly benefit from user inputs.
Contributions can range from sending new data for their integration in
the database to code development.

\section*{Acknowledgement}
The authors thank Bjorn O. Mysen (Carnegie
Institution for Science), Tobias K. Bechgaard (Novo Nordisk), Lothar Wondraczek (University of Erlangen),
and Daniel R. Neuville (CNRS-IPGP) for the provision of raw data for
their inclusion in the database. CLL thanks Andrew Valentine (Durham
University) for various discussions and advice regarding optimization
and machine learning.  Constructive comments from anonymous reviewers were highly appreciated.  Numerical computations were performed on the
S-CAPAD plateforme, IPGP, France.

\section*{Funding}
CLL acknowledges funding from a Chaire d’Excellence from the IdEX Université de Paris ANR-18-IDEX-0001. CLL and BB acknowledge funding from the Data intelligence institute of Paris, IdEX Université de Paris ANR-18-IDEX-0001.

\section*{Author contributions}
Charles Le Losq: Conceptualization, Methodology, Software, Validation, Formal analysis, Investigation, Resources, Data Curation, Writing - Original Draft, Writing - Review \& Editing, Visualization, Supervision, Funding acquisition, Project administration.

Barbara Baldoni:  Methodology, Software,  Formal analysis, Investigation, Data Curation,  Writing - Review \& Editing, Visualization

\section*{Competing interests}
Authors declare no competing interests. 

\section*{Materials \& Correspondence}
All the data are available in the main text or the supplementary materials. The database and the computer code to reproduce the results of this study is available as a Python library at the web address \url{https://github.com/charlesll/i-melt} and on Zenodo \cite{lelosq2023b}.  Correspondence can be addressed to the corresponding author.

\bibliographystyle{ieeetr}  
\bibliography{ms}  

\end{document}